\documentclass[12pt]{article}
\usepackage{amsmath}
\usepackage{amssymb}
\usepackage{amsfonts}
\catcode`\@=11
\catcode`\@12
\def\super       {\hat{sl}(2|1;\mathbb C)_k}
\def\iso         {\hat{sl}(2;\mathbb C)}

\def\N           {\mathbb N}

\def\Z           {\mathbb Z}
\def\R           {\mathbb R}
\def\C           {\mathbb C}
\def\ben         {\begin{equation}}
\def\een         {\end{equation}}
\def\bea         {\begin{eqnarray}}
\def\eea         {\end{eqnarray}}
\def\nn          {\nonumber \\ }

\def\hslc        {\hat{sl}(2|1;{\mathbb C})}
\def\hslck       {\hat{sl}(2|1;{\mathbb C})_k}

\def\hf          {\frac{1}{2}}

\begin{document}
\def\theequation{\arabic{equation}}
\begin{titlepage}
\begin{flushright}
DTP/98/5\\
March 1998\\
\end{flushright}
\vspace{1cm}  
\begin{center}
{\Large\bf Parafermionic representation of the affine $\hslc$
algebra at fractional level.}\\
\vspace{1cm}
{\large P. ~Bowcock, M. ~Hayes and 
A. ~Taormina}\\
\vspace{0.5cm}
{\it Department of Mathematical Sciences, University of Durham, Durham, DH1 ~3LE, England}\\
\vspace{3cm}
\end{center}
%
%
%
%
\begin{abstract}
The four fermionic currents of the affine superalgebra $\hslc$ at fractional level $k=\frac{1}{u}-1, u \in \N $ are shown to be realised in terms of
a free scalar field, an $\iso$ doublet field and a primary field of the
parafermionic algebra $\Z_{u-1}$.  
\end{abstract}
\vskip 8truecm
{}e-mail: Peter.Bowcock@durham.ac.uk, M.R.Hayes@durham.ac.uk, 
Anne.Taormina@durham.ac.uk
\end{titlepage}
Affine Lie algebras and superalgebras at fractional level have received
some attention in the last years in the context of the study of non critical strings \cite{yankiel,fy1,huyu}. Our original motivation in the analysis 
of the representation theory of $\hslc$ at fractional level $k$ is its 
potential relevance in the description of $N=2$ (non critical) superstrings.
We take the matter coupled to supergravity in an $N=2$ super Coulomb gas
representation with central charge
\ben
c_{\text{matter}}=3(1-\frac{2p}{u}),~~~~p, u \in \N,~~~{\rm gcd}(p,u)=1.
\end{equation}
Then, the level $k$ of the matter affine superalgebra $\hat{sl}(2|1)$ appearing in the 
\newline
$SL(2|1;\R)/SL(2|1;\R)$ gauged Wess-Zumino-Novikov-Witten model, believed to be intimately related
to the $N=2$ string, is of the form
\ben
k = \frac{p}{u} - 1.
\label{levelp}
\end{equation}
This is precisely the type of fractional levels first discussed in the paper
by Kac and Wakimoto on admissible representations of affine Lie algebras
\cite{KW88}. The detailed study of the representation theory of $\hslc$ 
for levels of the form \eqref{levelp} with $p=1$ (which leads to the
description of {\em unitary} minimal $N=2$ matter) has revealed the
existence of a finite number of irreducible representations
whose characters form a finite representation of the modular group \cite{HT98}. There are thus {\em rational}, although non unitary, theories associated with $\hslc$ at fractional level.

One of the very interesting byproducts of the above analysis is the existence of
a representation of the $\hslc$ currents in terms of $\Z _ {u-1}$
parafermions, as we shall now describe.

At level $k=\frac{1}{u}-1$, the vacuum representation is in the Ramond sector
of the theory and is labeled by the two quantum numbers $h_-^R=0$ and $h_+^R=0$,
which correspond to the isospin and $U(1)$ charges of the highest weight state
\cite{BT96,BHT97}. The associated character is given by,
\begin{gather}
\chi^{R,IV,\super}_{0,0}(\tau,\sigma,\nu)=
F^R(\tau,\sigma, \nu)\times \notag\\[3mm]
\sum_{a\in\Z}q^{a^{2}pu}z^{-ap}\frac{1-q^{2ua}z^{-1}}{
(1+q^{au}z^{-\frac{1}{2}}\zeta^{-\frac{1}{2}})(1+q^{au}z^{-\frac{1}{2}}
\zeta^{\frac{1}{2}})},\label{vacchar}
\end{gather}
where $q=e^{2i\pi \tau},~z=e^{2i\pi \sigma},~\zeta=e^{2i\pi \nu}$
and the factor $F^R(\tau,\sigma,\nu)$ is,
\begin{equation}
F^{R}(\tau,\sigma, \nu) =
\prod^{\infty}_{n=1}\frac{(1+z^{\frac{1}{2}}\zeta^{\frac{1}{2}}q^{n})
(1+z^{-\frac{1}{2}}\zeta^{\frac{1}{2}}q^{n-1})
(1+z^{\frac{1}{2}}\zeta^{-\frac{1}{2}}q^{n})
(1+z^{-\frac{1}{2}}\zeta^{-\frac{1}{2}}q^{n-1})}
{(1-q^{n})^{2}(1-zq^{n})(1-z^{-1}q^{n-1})}.
\label{factorR}
\end{equation}
The crucial ingredient needed for the construction of the new representation
of $\hslc$ currents is provided by the decomposition of the above character
in characters of the affine $\iso$ subalgebra at the same fractional level. The general decomposition
formulae appear in \cite{HT98} for the Neveu-Schwarz sector of the theory. 
Using spectral flow as discussed in \cite{BHT97}, and the following
relation between
the $\iso$ characters of isospin $j(n,n')=\hf (n-n'(k+2))$ and isospin
$j(u-n-1,u-n'-1)$,
\ben
\chi^{\iso_k}_{n,n'}(\tau, -\sigma-\tau)
=z^{-\frac{k}{2}}q^{-\frac{k}{4}} 
\chi^{\iso_k}_{u-n-1,u-n'-1}(\tau,\sigma ),
\end{equation}
one easily  obtains the Ramond vacuum decomposition,
\begin{eqnarray}
&&\chi^{R,IV,\super}_{0,0}(\tau,\sigma,\nu)=\nn
&&\sum_{i=0}^{u-1} \sum_{s=0}^{u-2} c_{i,-3i-2us}^{(u-1)}(\tau)
\theta_{u(u-1)(i+1)+2iu(\frac{u}{2}-[\frac{u}{2}])-2us,u(u-1)}(\tau, \frac{\nu}{u}) \chi_{u-1-i,0}^{\iso_k}(\tau,\sigma),\nn
\label{decomp}\end{eqnarray}
where the symbol $[\frac{u}{2}]$ is the integer part of $\frac{u}{2}$. 
The above formula provides some insight into the coset structure
$\frac{\hslck}{\iso_k}$. Indeed,
the functions $c_{\ell,m}^{(u-1)}(\tau)$ are the $\hat{su}(2)_{u-1}$
string functions introduced in \cite{KP}. When multiplied by the Dedekind
function $\eta(\tau)$, they provide the partition functions of the 
parafermionic algebra  $\Z _{u-1}$
whose lowest dimensional fields are labeled $\Phi^{\ell}_{m}$ \cite{ZAM,Qiu,GepQiu}.
 Furthermore, the theta functions at level $u(u-1)$, when
divided by the Dedekind function, are the characters for the rational
torus $A_{u(u-1)}$ \cite{DVV}. The coset structure is also reflected in the
following relations between central charges.
The central charge of the Virasoro algebra associated to $\hslck$ is,
\ben
c_{\hat{sl}(2/1)}=\frac{k~{\rm sdim~} sl(2/1)}{k+1}=0
\end{equation}
since there is an equal number of bosonic and fermionic generators in 
$sl(2/1)$. As a consequence,
the central charge of the coset $\frac{\hslck}{\iso_k}$, for $k=\frac{1}{u}-1$, is
\ben
c_{{\rm coset}}=-c_{{\rm sl(2)}}=\frac{3(u-1)}{u+1}.
\end{equation}
It can in turn be rewritten as,
\ben
c_{{\rm coset}}=c_{{\rm torus}}+c_{{\rm paraf}}=1+\frac{2(u-2)}{u+1}.
\end{equation}

A close reading of the decomposition formula \eqref{decomp} provides the
information needed to construct the new $\hslck$ representation.
By expanding the vacuum character \eqref{vacchar} in powers of $q$, one discovers at grade one the eight states belonging to the
adjoint representation of 
$sl(2/1;\C)$. It is also straightforward to identify on the right hand side
of the decomposition \eqref{decomp}, the four states at grade one which are neutral with respect to the $U(1)$ algebra, i.e. those which are independent of the variable $\zeta$. They all
lie in the term $i=u-1, ~s=0$ and correspond to the first bosonic excitations of the vacuum. The other four states correspond to the first
fermionic vacuum excitations and lie in the terms $i=u-2,~s=[\frac{u}{2}],
s=[\frac{u}{2}]-1$. The string function appearing in these two terms is
$c_{1,1}^{(u-1)}(\tau)$, which is, upon multiplication by the Dedekind function, the parafermionic partition function 
with lowest dimensional field $\Phi^1_1(z)$, identified with the primary field $\sigma_{(1)}(z)$ with conformal
dimension \cite{ZAM},
\ben
h_{(1)}=\frac{u-2}{2(u-1)(u+1)}.
\end{equation}     
This field is one of the $(u-2)$ conformal spin fields or order parameters
associated with the parafermion algebra based on $\Z_{u-1}$ \cite{ZAM}.
It enters in our representation of the fermionic $\hslc$ currents as follows.
Adopting the notations introduced in \cite{BT96} for the $\hslc$ currents,
one writes,
\begin{eqnarray}
j^{\pm}(z)&=&
\Delta ^{\pm}(z)(i\sqrt{\frac{1-u}{u}}):e^{\mp \frac{i}{2}\sqrt{\frac{2u}{u-1}}\phi(z)}:\sigma_{(1)}(z),\nn
j^{\pm ' }(z)&=&
\Delta ^{\pm}(z)(i\sqrt{\frac{1-u}{u}}):e^{\pm \frac{i}{2}\sqrt{\frac{2u}{u-1}}\phi(z)}:\sigma_{(1)}(z),
\label{fermcur}\end{eqnarray}
where $\Delta^+(z)$ is the field corresponding to the $\iso_k$ doublet
highest weight state with conformal weight $h_{{\rm doublet}}=\frac{3u}{4(u+1)}$ and O.P.E. \cite{AY},
\begin{eqnarray}
\Delta^+(z)\Delta^-(w)&=& (z-w)^{-\frac{3u}{2(u+1)}}+\frac{u}{1-u}(z-w)^{-\frac{u-2}{2(u+1)}}J_3(w)+O((z-w)^
{\frac{u}{2(u+1)}}),\nn
\Delta^{\pm}(z)\Delta^{\pm}(w)&=& 
\frac{u}{1-u} (z-w)^{-\frac{u-2}{2(u+1)}}J^{\pm}(w)+O((z-w)^
{\frac{u}{2(u+1)}}),
\end{eqnarray}
where $J^{\pm}(\omega)$ and $J^3(\omega)$ are the $\iso$ currents. 
As usual, the vertex operators have conformal weight $h_{U(1)}=\frac{u}{4(u-1)}$ and obey the O.P.E.,
\ben
:e^{\pm \frac{i}{2}\sqrt{\frac{2u}{u-1}}\phi(z)}:
:e^{\pm \frac{i}{2}\sqrt{\frac{2u}{u-1}}\phi(w)}:~=
(z-w)^{\frac{u}{2(u-1)}}:e^{\pm \frac{ i}{2}\sqrt{\frac{2u}{u-1}}(\phi(z)
+\phi(w))}:
\label{ope1}\end{equation}
and
\ben
:e^{\pm \frac{i}{2}\sqrt{\frac{2u}{u-1}}\phi(z)}:
:e^{\mp \frac{i}{2}\sqrt{\frac{2u}{1-u}}\phi(w)}:~=
(z-w)^{-\frac{u}{2(u-1)}}:e^{\pm \frac{ i}{2}\sqrt{\frac{2u}{u-1}}(\phi(z)
-\phi(w))}:~.
\label{ope2}\end{equation}
Note that 
\ben
h_{{\rm doublet}}+h_{U(1)}+h_{(1)}=1,
\end{equation}
as required for the fermionic $\hslck$ currents.
So, the parafermionic representation of $\hslc$ we obtained is in terms of
the $\iso$ currents, the $U(1)$ current
\ben
U(z)=i\sqrt{\frac{u-1}{2u}}\partial \phi(z),
\label{u}\end{equation}
where $\phi(z)$ is a free scalar field with the standard O.P.E.
\ben
\phi(z)\phi(w) \sim -{\rm ln}|z-w|,
\end{equation}
and the four fermionic currents \eqref{fermcur}. Note that
the parafermions enter in this representation in a rather different way
than in representations of $\iso$ at {\em integer} level $k$ (see e.g. \cite{HNY}),
where the parafermionic algebra is $\Z _k$.

We now illustrate the case where $u=3$, when the parafermions are the 
usual fermions. The spin field of interest is then $\sigma(z)$
of weight $1/16$, and appears in the description of the Ising model. The relevant O.P.E. are,
\begin{eqnarray}
&&\sigma(z)\sigma(w)=(z-w)^{-1/8}+\hf (z-w)^{3/8}\psi(w) +...\nn
&&\Delta^+(z) \Delta^-(w) = (z-w)^{-9/8} -\frac{3}{2}(z-w)^{-1/8}J^3(w)
+\alpha (z-w)^{3/8} A_3(w)+...\nn
&&\Delta^{\pm}(z)\Delta^{\pm}(w)=-\frac{3}{2}(z-w)^{-1/8}J^{\pm}(w)+
\alpha'(z-w)^{3/8}A^{\pm}(w)+...,
\label{ope3}\end{eqnarray}
where $\psi(w)$ is a fermion and $A^{\pm}(w),A_3(w)$ form an $\iso$ triplet
of weight $\frac{3}{2}$. These fields, whose occurrence in the theory can be
traced back to the vacuum character formula \eqref{decomp}, have to be taken into account in parafermionic representations of normal ordered products
of currents, but are not relevant to the discussion of commutation relations
of those currents. The exact value of the two parameters $\alpha $ and $\alpha'$ in the above O.P.E. is therefore not required for our purpose.
Now, the short distance behaviour of $\hslck$ currents is given by,
\begin{eqnarray}
&&j^+(z)j^-(w) \sim [U(w)+J^3(w)] (z-w)^{-1} +k (z-w)^{-2},\nn
&&j^{+'}(z)j^{-'}(w) \sim [U(w)-J^3(w)] (z-w)^{-1}-k(z-w)^{-2},\nn
&&j^{\pm '}(z)j^{\pm}(w) \sim J^{\pm}(w) (z-w)^{-1},\nn
&&U(z)U(w) \sim -\frac{k}{2} (z-w)^{-2},~~~~~
J^3(z)J^3(w) \sim \frac{k}{2} (z-w)^{-2},\nn
&&2J^3(z)j^{\pm}(w) \sim \pm j^{\pm}(w) (z-w)^{-1},~~~~~
2J^3(z)j^{\pm '}(w) \sim \pm j^{\pm '}(w) (z-w)^{-1},\nn
&&2U(z)j^{\pm}(w) \sim \mp j^{\pm}(w) (z-w)^{-1},~~~~~
2U(z)j^{\pm '}(w) \sim \pm j^{\pm '}(w) (z-w)^{-1},\nn
&&J^{\pm}(z)j^{\mp}(w) \sim \mp j^{\pm '}(w)(z-w)^{-1},~~~~~
J^{\pm}(z)j^{\mp '}(w) \sim \pm j^{\pm }(w)(z-w)^{-1},\nn
&&J^+(z)J^-(w) \sim 2J^3(w)(z-w)^{-1} + k (z-w)^{-2}.
\end{eqnarray}
Using \eqref{ope1} and \eqref{ope2} together with
the O.P.E. \eqref{ope3} truncated to the terms with negative powers of $(z-w)$, 
one checks that the fermionic currents as given in \eqref{fermcur} and the
bosonic current \eqref{u} provide, with the $\iso$ currents $J^{\pm}(z)$ and
$J^3(z)$, a representation of $\hslc$ at level $k=-2/3$.

A similar representation in the case $u=4$ involves the spin field of the
3-states Potts model, with conformal weight $1/15$. This is hardly surprising,
since the $\Z_2$ and $\Z_3$ \break
parafermionic theories coincide with the 
Ising and 3-states Potts models respectively. At the level of characters, one actually has,
\ben
\chi^{Vir(3)}_{1,1}(\tau)=\eta(\tau)c^{(2)}_{2,2}(\tau),~~
\chi^{Vir(3)}_{2,1}(\tau)=\eta(\tau)c^{(2)}_{0,2}(\tau),~~
\chi^{Vir(3)}_{2,2}(\tau)=\eta(\tau)c^{(2)}_{1,1}(\tau),
\end{equation}
and
\begin{gather}
\chi^{Vir(5)}_{1,1}(\tau)+\chi^{Vir(5)}_{4,1}(\tau)=
   \eta(\tau)c^{(3)}_{0,0}(\tau),~~\chi^{Vir(5)}_{4,3}(\tau)=
       \eta(\tau)c^{(3)}_{3,1}(\tau)\\
\chi^{Vir(5)}_{2,1}(\tau)+\chi^{Vir(5)}_{3,1}(\tau)=
   \eta(\tau)c^{(3)}_{2,0}(\tau),~~\chi^{Vir(5)}_{3,3}(\tau)=
       \eta(\tau)c^{(3)}_{1,1}(\tau).
\end{gather}

The deep r{\^o}le of this representation in the description of noncritical 
$N=2$ strings with minimal, unitary matter is yet to be uncovered. However,
it is worth noticing that there is a direct relation between the parafermionic
algebra involved in the representation of the $\hslck$ currents at level
$k=\frac{1}{u}-1$,
and the parafermionic algebra involved in the realisation of the $N=2$
superconformal algebra when the central charge is \eqref{levelp} with $p=1$. 
The former is $\Z_{u-1}$ while the latter is $\Z_{u-2}$.

\vskip 2cm

{\bf Acknowledgements}

We would like to thank H. Kausch for useful comments. M. Hayes acknowledges the British EPSRC for a studentship. A. Taormina acknowledges the Leverhulme Trust for a fellowship and thanks the EC for 
support under a Training and Mobility of Researchers Grant No. FMRX-CT-96-0012.

\def\NPB{Nucl.\ Phys.\ B} 
\def\PRD{Phys.\ Rev.\ D} 
\def\PLB{Phys.\ Lett.\ B} 
\def\MPLA{Mod.\ Phys.\ Lett.\ A}
\def\CMP{Commun.\ Math.\ Phys.}  
\def\IJMPA{Int.\ J.\ Mod.\ Phys.\ A}

\end{document}